\documentclass[sn-mathphys,Numbered]{sn-jnl}% Math and Physical Sciences Reference Style
\usepackage{graphicx}%
\usepackage{multirow}%
\usepackage{amsmath,amssymb,amsfonts}%
\usepackage{amsthm}%
\usepackage{mathrsfs}%
\usepackage[title]{appendix}%
\usepackage{xcolor}%
\usepackage{textcomp}%
\usepackage{manyfoot}%
\usepackage{booktabs}%
\usepackage{algorithm}%
\usepackage{algorithmicx}%
\usepackage{algpseudocode}%
\usepackage{listings}%
\theoremstyle{thmstyleone}%
\theoremstyle{thmstyletwo}%

\theoremstyle{thmstylethree}%

\begin{document}

\title[The Role of Soil Surface in a Sustainable Semiarid Ecosystem]{The Role of Soil Surface in a Sustainable Semiarid Ecosystem}


\author*[1]{\fnm{Swadesh} \sur{Pal}}\email{spal@wlu.ca}

\author[2]{\fnm{Malay} \sur{Banerjee}}\email{malayb@iitk.ac.in}

\author[1,3]{\fnm{Roderick} \sur{Melnik}}\email{rmelnik@wlu.ca}

\affil[1]{\orgdiv{MS2Discovery Interdisciplinary Research Institute}, \orgname{Wilfrid Laurier University}, \orgaddress{\city{Waterloo}, \postcode{N2L3C5}, \state{Ontario}, \country{Canada}}}

\affil[2]{\orgdiv{Department of Mathematics and Statistics}, \orgname{IIT Kanpur}, \orgaddress{\city{Kanpur}, \postcode{208016}, \state{Uttar Pradesh}, \country{India}}}

\affil[3]{\orgdiv{BCAM - Basque Center for Applied Mathematics}, \orgname{Organization}, \orgaddress{\city{Bilbao}, \postcode{E-48009}, \state{Biscay}, \country{Spain}}}


\abstract{
Patterns in a semiarid ecosystem are important because they directly and indirectly affect ecological processes, biodiversity, and ecosystem resilience. Understanding the causes and effects of these patterns is critical for long-term land surface management and conservation efforts in semiarid regions, which are especially sensitive to climate change and human-caused disturbances. It is known that there is a regular connection between the vegetation and the living species in a habitat since some animals evolved to live in a semiarid ecosystem and rely on plants for food. In this work, we have constructed a coupled mathematical model to connect the water resource, vegetation and living organisms and have investigated how the soil surface affects the resulting patterns for the long term. This study contributes to a better understanding of ecological patterns and processes in semiarid environments by shedding light on the complex interaction mechanisms that depend on the structure of semiarid ecosystems. The findings provide further critical insight into the influence of efforts for improving ecosystem resilience and adjusting to the challenges posed by climate change and human activities.
}




\maketitle

\section*{Introduction}\label{sec1}

Vegetation is an important component of an ecosystem because it helps energy and nutrient cycles, improves air quality, reduces flooding and soil erosion, and many other mechanisms \cite{simonich1994,berdugo2020}. Vegetation survives only on sunlight and groundwater minerals from the soil. In general, the availability of sunlight is not a problem for the vegetation, but sometimes the groundwater levels cause problems in semiarid ecosystems due to irregular rainfall \cite{birkett2005}. This causes further food limitations for vegetation-dependent herbivores because they maintain a healthy balance by preventing vegetation's overgrowth \cite{braun2021}. It is known that diversity can be preserved through life-history trade-offs between growth rate and competitive ability \cite{anand2010}. This is the prediction of the intermediate disturbance hypothesis, which was first developed as a spatially homogeneous theory. Unlike theories that presume 'well-mixed' disturbance, patch dynamics predicts species diversity in relation to the intensity and magnitude of newly generated gaps \cite{huston1979}. In addition, many plants rely on herbivores, such as bees, to help them in pollination, and they enhance plant biodiversity at high productivity but have the reverse impact at low productivity \cite{koerner2018}. As a result, low rainfall has a significant effect on regulating several food chains and food webs.

Researchers have been using mathematical models as tools to capture and predict their behaviours in the long term. For instance, pattern formation in ecology helps in understanding the evolution of various living organisms in conjunction with non-living components. It is one of the key prerequisites that allows differently adapted species to coexist on a regional scale \cite{amarasekare2003}. Many ecological theories have been developed to understand complex ecological systems and their interactions based on heterogeneous stationary and dynamic patterns \cite{murray2001}. Moreover, some important developments have been taking place in the context of complex systems that directly impact complex ecological systems studies \cite{levin2005,tadic2021}. Among others, research has also been progressing on vegetation patterns under climate change and water cycle pathways and their related factors, such as land surface coupled to other components of climate, through coupled climate models reflecting these systems' complexity \cite{mihailovic1997,melnik1998,dirmeyer2014,sun2022}. Rich dynamics and associated patterns of different components of climate systems, including water and soil, have also been analyzed with a wide range of simplified mathematical models \cite{wang2017, sardanyes2018, rajagopal2020, sun2022a, lei2022}. In general, the state of an ecosystem responds smoothly to continuous trends to a gradual change in external conditions, such as climate, groundwater reduction, habitat fragmentation, and many more \cite{vitousek1997,tilman2001,scheffer2001}. But sometimes, a crucial situation arises when a drastic change happens in an ecosystem where the reversal is almost impossible. Such situations are often studied in the context of tipping phenomena and points of no return in ecosystems (e.g., \cite{o2020} and references therein). There may be some indications or even no warning at all behind such drastic change \cite{scheffer2001,scheffer2003}. This sudden dramatic shift of nature is called a catastrophic shift, and the critical threshold is known as `catastrophic bifurcation' \cite{muratori1989, kuznetsov1998}. 

Some examples of this type of drastic change include the invasion of new species, a variety of human interventions, lightning-caused wildfires, etc. \cite{scheffer2009,saco2018}. Heavy rain or extreme drought can also contribute to such climate events \cite{sun2022}. Therefore, preventing catastrophic change before its occurrence is essential for long-term ecosystem functioning. The major challenges to understanding such dramatic change are inefficient data and the infancy of mathematical theory. A number of indicators have been proposed prior to this shift \cite{scheffer2009}, but to the best of our knowledge, all the underlying mathematical theory has been discussed only through saddle-node bifurcation \cite{hastings2018}; however, this article is intended to explain a new underlying mechanism (transcritical bifurcation) for the regime shift in the context of vegetation pattern.

Several mathematical models show long transient behaviours before converging to the final attractor \cite{hastings2018}, and they do not produce any clear trend in their properties. These long transients generally occur in a system close to different local and global bifurcation thresholds \cite{kuznetsov1998}. Sometimes, they do not converge to long-term stable solutions like constant or cyclic dynamics but persist for as many as tens of generations. In addition, a cyclic synchronous ecological system may lead to extinctions in the long run due to unexpected population fluctuations. For instance, a disturbance in two predators and one prey synchronously fluctuating population model may experience an overall predator populations boom, leading to a decrease in the prey populations, resulting in a sudden drop in the predator populations, and acting as a ripple effect throughout the ecosystem. On the other hand, long transients in ecological systems maintain biodiversity, making the ecosystem resilient and sustaining itself in the longer run \cite{cadotte2011}. Therefore, we intended to study long transients in the context of vegetation patterns. 

The soil surface of the earth is not uniform for all locations, and vegetation depends on it. Generally, the soil water travels through the ground in two ways: flow over the soil surface and underground diffusion \cite{deblauwe2008}, characterised by the soil surface. In this work, we have added water diffusion into the existing model of water-plant dynamics in a semiarid ecosystem \cite{klausmeier1999} and studied the role of the soil surface in a semiarid ecosystem. In addition, various animals and insects adapted to live in semiarid ecosystems; they depend on plants as their food source. Therefore, it is interesting to see how introducing herbivore species in the modified water-plant model affects the overall system's dynamics and enables us to understand the ecosystem better in the context of sustainability. 

\section*{Results}{\label{sec5}}

As a baseline, we first consider the scenario where herbivore species are absent and study water-plant dynamics with and without water flow and its diffusion. Then, we discuss the effect of water flow and water diffusion in the presence of herbivores.

\subsection*{Absence of herbivores}

As discussed, the temporal model without herbivores has at most three equilibrium points: water-only equilibrium $(u_{0},v_{0})$, which exists and is linearly stable for all the parameter conditions, and two coexisting equilibrium points can originate through a saddle-node bifurcation, and they are $(u_{1},v_{1})$ and $(u_{2},v_{2})$. These two coexisting equilibrium points exist for $\alpha > 2\beta$, and it is true for the parameter values given in Table \ref{tab1}. The linear stability analysis shows that one of them, $(u_{1},v_{1})$, is stable, and the other $(u_{2},v_{2})$ is unstable. Therefore, the temporal model is bistable; one is vegetated, and the other is bare. Water and plants cohabit in both circumstances, yet a minor perturbation in their unstable coexistence can result in the loss of their coexistence; however, they can maintain stable coexistence under small fluctuations. From an ecological perspective, bistability's underlying dynamical structure leads to the possibility that even minor perturbations - such as those resulting from climactic events, natural fluctuations, or interactions with humans - can cause profound changes in ecosystems. These can lead to persistent switches between vegetation and bare. Furthermore, the numerical simulations suggest that the basin of attraction for the bare equilibrium point decreases with increasing water supply ($\alpha$), and the reverse situation happens for the vegetated equilibrium point. This shows that the vegetation's extinction probability is low for higher rainfall. 

Vegetation growth and spatial distribution are influenced by the climate, the physical environment, and human activity \cite{sun2022}. Many mathematical models predict such patterns in the presence of species mobility.  For instance, a linearly stable coexisting equilibrium point for a temporal model may be unstable under a heterogeneous perturbation in the presence of diffusion terms \cite{turing1990}. In \cite{klausmeier1999}, the author has considered the water flow along the $x$-direction and the diffusion present only in the plant equation. In this case, the Turing instability analysis shows that the stable coexisting steady-state remains stable under heterogeneous perturbation when the water flow ($a_{x}$) is absent \cite{turing1990}, and also for the flow up to a certain level (suppose $a_{x}^{c}$), which depends on the values of $\alpha$ and $\beta$. Therefore, the regular pattern formation is impossible on flat ground ($a_{x}=0$) and for the amount of water flows less than $a_{x}^{c}$. But, above the critical threshold of water flow $a_{x}^{c}$, the coexisting steady-state becomes unstable, and the system produces irregular transient solutions initially due to the heterogeneous perturbation of the initial conditions, and vegetation takes time to spread over the landscape with moving along the downhill. At a large time, the irregular transient solution changes into regular waves, and the vegetation patches keep the same distance between them. However, in the presence of water diffusion, the system produces Turing patterns for some diffusion coefficient irrespective of the water flow, as discussed in the earlier section. Furthermore, sufficient water flow induces stationary patterns to form strips propagating through the spatial domain.

\subsection*{Herbivores' influence}

\begin{figure}[ht!]
\centering
\includegraphics[width=0.85\textwidth]{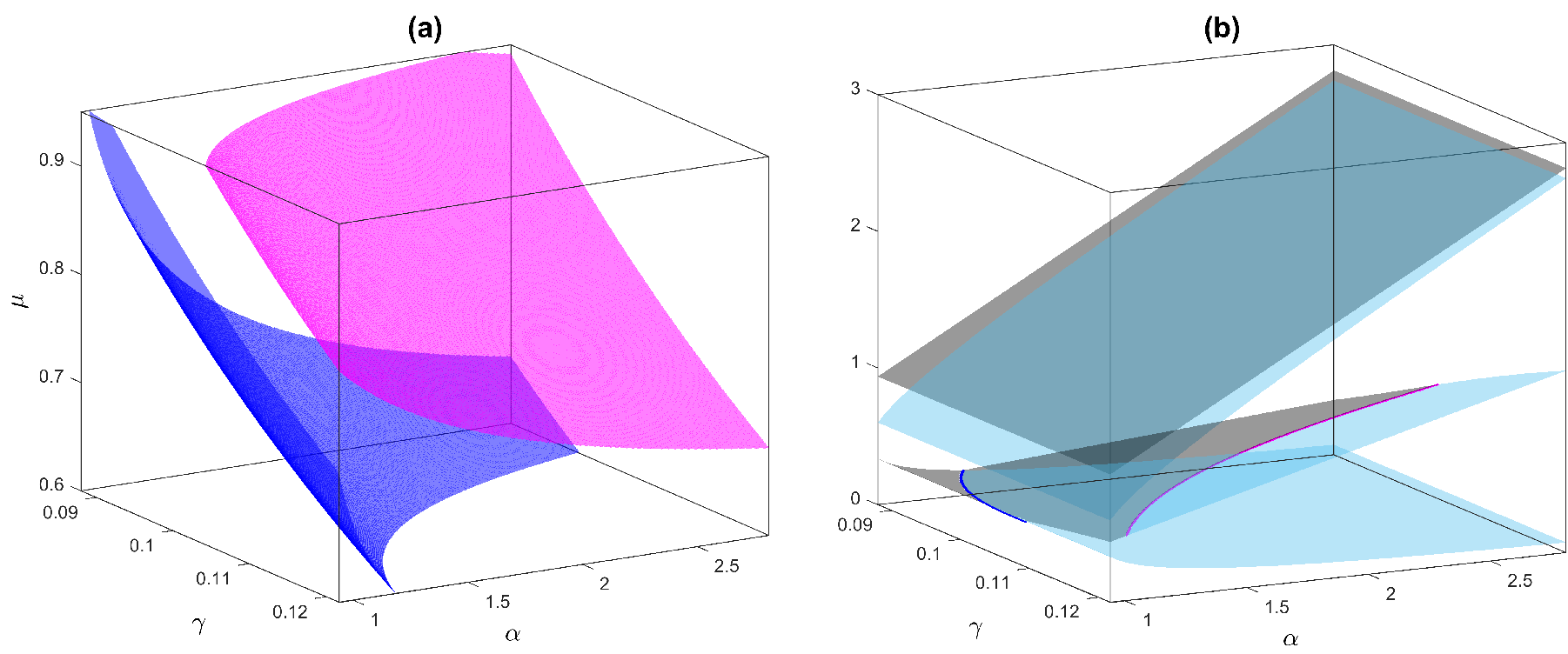}
\caption{(Colour online) (a) Three-parametric bifurcation diagram for the non-trivial equilibrium point for the temporal model, and (b) the possible equilibrium points corresponding to the water density ($u$) for the bifurcation parameters $\alpha$ and $\gamma$ with a fixed value $\mu = 0.8$. The blue and magenta colour surfaces (curves) represent the transcritical and Hopf bifurcations, respectively. The grey and cyan colour surfaces represent the stable and unstable equilibrium points. The fixed parameter values for both figures are $\beta = 0.45, h = 1$, and $\eta = 0.05$.}\label{Fig1}
\end{figure}

Diversity is an essential component of a system's persistence and adaptability in the face of environmental change \cite{anand2010}, and here we discuss it by introducing the herbivores' interference into the existing water-plant model. After introducing the herbivores' dynamics, the resulting three-species model preserves all its equilibrium points with zero herbivores as the extra component. It has the water-only steady-state $(u_{0},v_{0},w_{0}) = (\alpha,0,0)$, which is linearly stable. The steady-states $(u_{1},v_{1},0)$ and $(u_{2},v_{2},0)$ exist for $\alpha>2\beta$, and $(u_{2},v_{2},0)$ is always unstable. The stability of $(u_{1},v_{1},0)$ depends on the feasible existence of the coexisting equilibrium point  $(u_{*},v_{*},w_{*})$. This coexisting equilibrium point bifurcates through a transcritical bifurcation at $(u_{1},v_{1},0)$ [see the blue surface in Fig. \ref{Fig1}(a)], and the coexisting equilibrium point gains stability and remains stable between blue and magenta surfaces. It loses stability through a sub-critical Hopf bifurcation (magenta surface in Fig. \ref{Fig1}(a)), and the water-only state remains the only linearly stable state. As the system is bistable, a trajectory can go to either of these equilibrium points depending on the initial conditions belonging to the appropriate basins of attraction.

\begin{figure}[ht!]%
\centering
\includegraphics[width=0.95\textwidth]{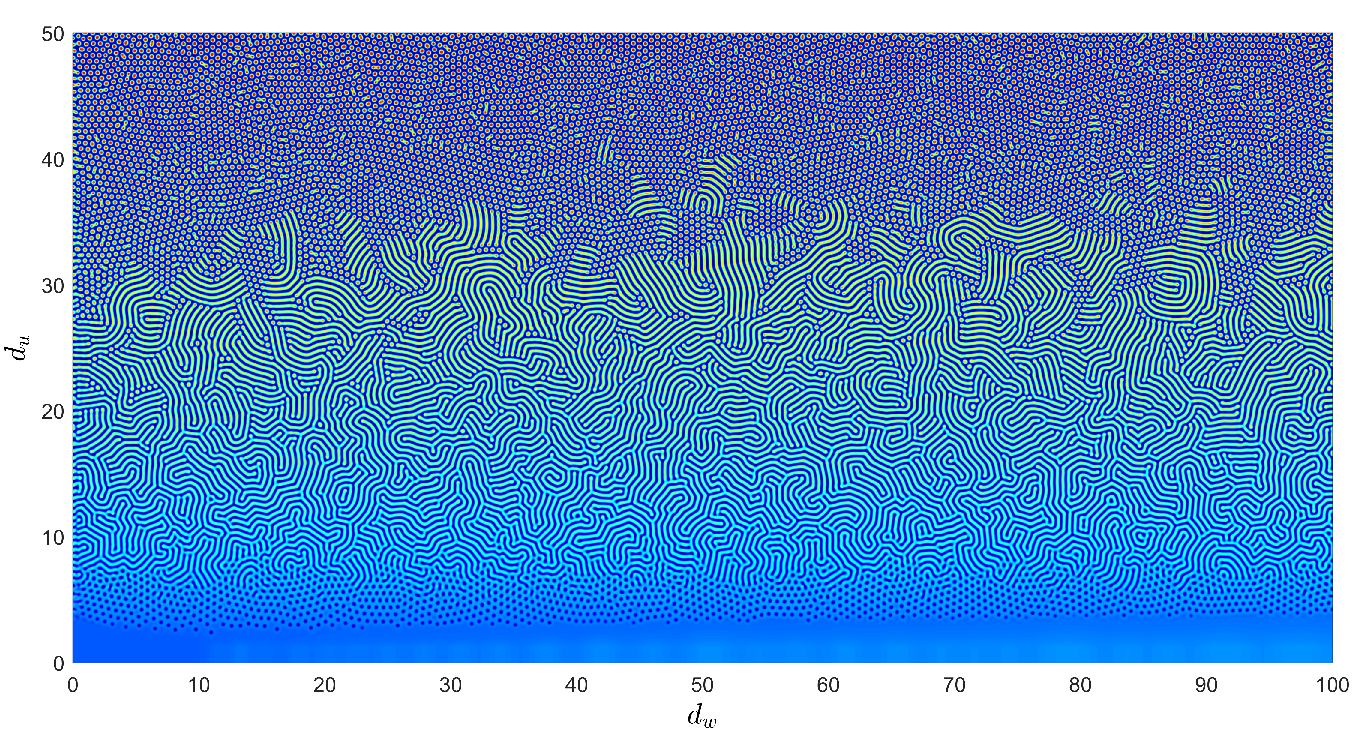}
\caption{ The patterns for the plants in the absence of downhill water flow for various combinations of diffusion rates for water and herbivores. The fixed parameter values are: $\alpha = 2.8125$, $\beta = 0.45$, $\gamma = 0.1$, $h = 1$, $\mu = 0.8$, and $\eta = 0.05$.}\label{Fig2}
\end{figure}

\subsubsection*{Effect of groundwater diffusion} 

We first study the spatio-temporal dynamics of the model when water diffusion is present. We fix the temporal parameter values $\alpha = 2.8125, \mu  = 0.8$, and $\eta = 0.05$ and take $\gamma$ as the bifurcation parameter. The system produces a heterogeneous distribution for three components for $\gamma = 0.1$ for different values of $d_{u}$ and $d_{w}$. Figure \ref{Fig2} depicts the schematic diagram of patterns and the homogeneous solutions for the plants ($v$) for different combinations of $d_{u}$ and $d_{w}$, which matches with theoretical investigations. Here, the silico trials show that the hot-spot pattern for the water corresponds to the cold-spot patterns for the plants and herbivores. In reality, the water negatively correlates with plants and herbivores because excess water damages the vegetation. 

\begin{figure}[ht!]%
\centering
\includegraphics[width=\textwidth]{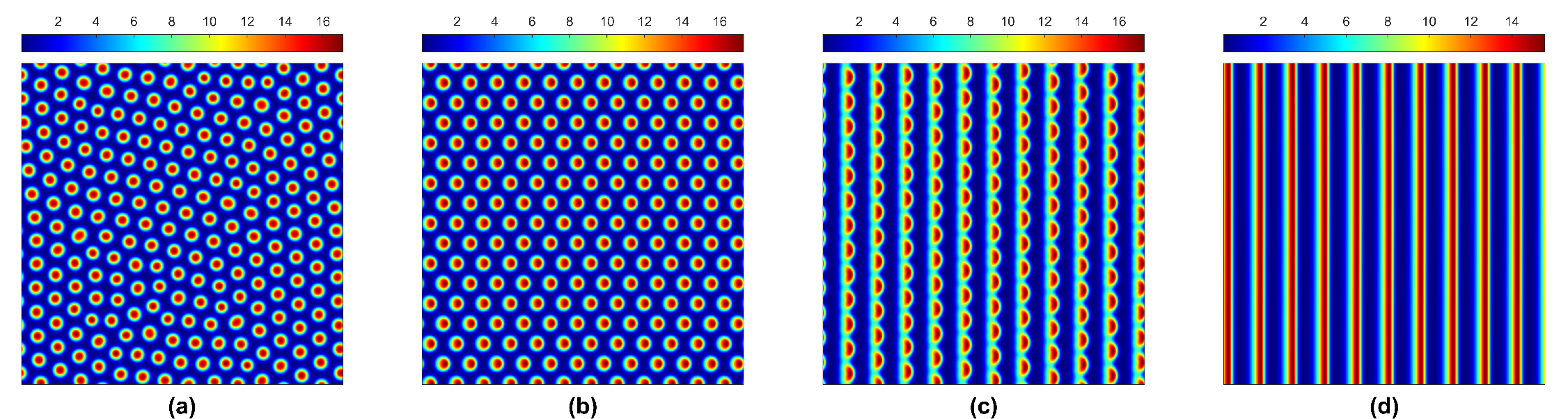}
\caption{ (Colour online) Stationary and non-stationary solutions for the plant. Parameter values: $\alpha = 2.8125$, $\beta = 0.45$, $\gamma = 0.1$, $h = 1$, $\mu = 0.8$, $\eta = 0.05$, $a_{y} = 0$, $d_{u} = 500$, $d_{w} = 100$: (a) $a_{x} = 0$, (b) $a_{x} = 100$, (c) $a_{x} = 150$, and (d) $a_{x} = 182.5$.}\label{Fig3}
\end{figure}

The groundwater diffusion is generally low as it moves through the porous spaces between particles of unconsolidated soil or through networks of fractures and solution openings in consolidated rocks, which varies with locations \cite{harter2003}. Here, cold-spot patterns exist for the lower water diffusion on flat ground, and labyrinthine and hot-spot patterns form for higher diffusion rates [see Fig. \ref{Fig2}]. In addition, negative correlations are also observed between the organisms. For zero groundwater diffusion, the non-trivial homogeneous steady state is linearly stable for the temporal parameter values used in Fig. \ref{Fig2}, and no stationary pattern can be obtained [see Fig. \ref{Fig2}]. The coexisting steady-state satisfies the Turing instability conditions when the water diffusion parameter crosses the critical Turing threshold \cite{turing1990}, resulting in Turing patterns, e.g., see Fig. \ref{Fig3}(a). Furthermore, this critical diffusion threshold can be obtained for the water by considering zero diffusion for the herbivores, as observed in Fig. \ref{Fig2}. However, the critical diffusion threshold cannot be obtained for herbivores by taking zero diffusion for the water. This shows groundwater diffusion is crucial in forming stationary patterns on flat ground.

\subsubsection*{Effect of groundwater flow} 
\begin{figure}[h]%
\centering
\includegraphics[width=0.85\textwidth]{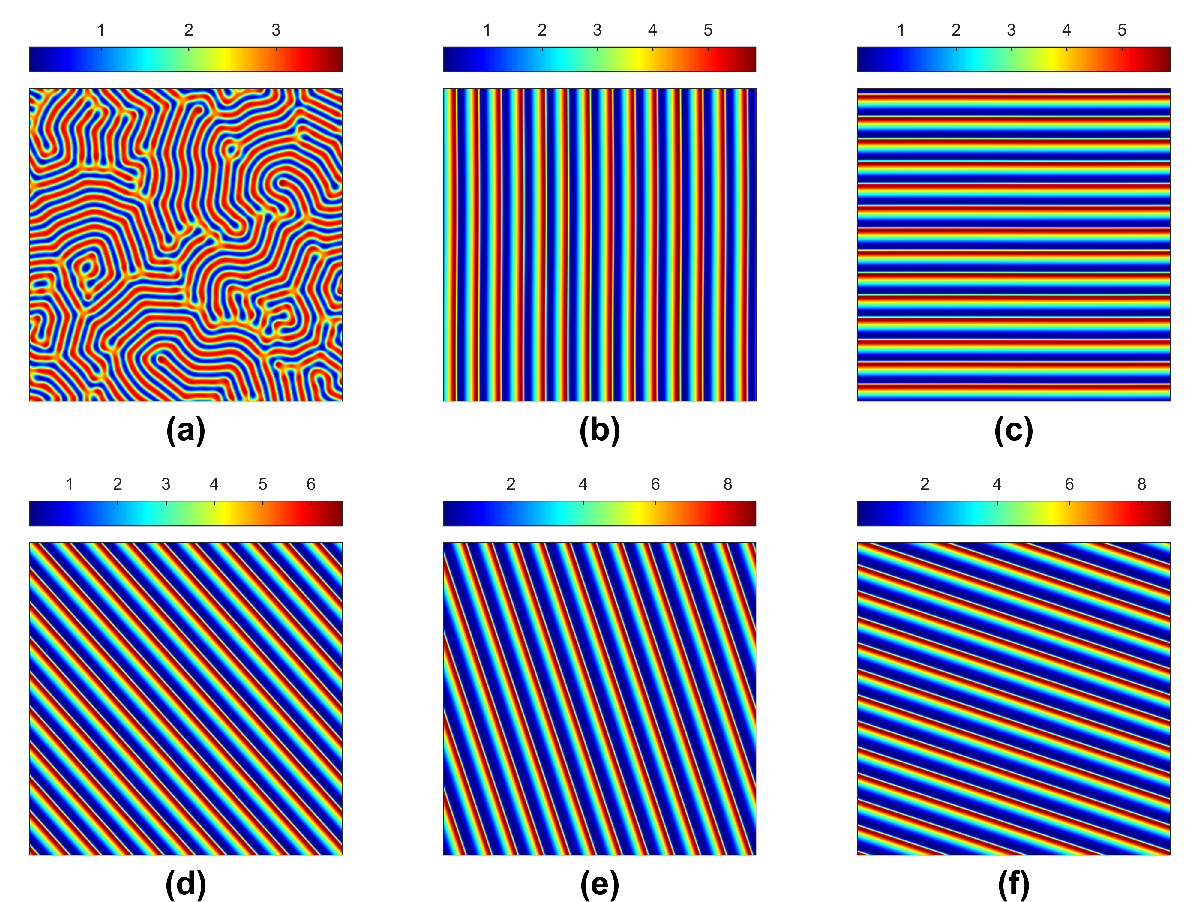}
\caption{ (Colour online) Stationary and non-stationary solutions for the plant for different values of the groundwater flow along with directions. Parameter values: $\alpha = 2.8125$, $\beta = 0.45$, $\gamma = 0.1$, $h = 1$, $\mu = 0.8$, $\eta = 0.05$, $d_{u} = 10$, $d_{w} = 20$: (a) $a_{x} = a_{y} = 0$, (b) $a_{x} = 10, a_{y} = 0$, (c) $a_{x} = 0, a_{y} = 10$, (d) $a_{x} = 10, a_{y} = 10$, (e) $a_{x} = 30, a_{y} = 10$, and (f) $a_{x} = 10, a_{y} = 30$.}\label{Fig4}
\end{figure}

The same type of moving strips for the water-plant model is observed for the herbivore-included model without groundwater diffusion. Here, the groundwater flow makes the homogeneous steady-state unstable but requires a minimum speed. Though this critical speed depends on the other parameter values, the speed variation is slight compared to other parameter changes. In addition, these regular stripes can move diagonally for considering the same amount of water flow along $x$ and $y$ directions [e.g., see Fig. \ref{Fig4}]. The ratio of water flow in the $x$ and $y$ directions determines the angle of moving regular strips. 

The non-homogeneous stationary patterns in Fig. \ref{Fig3}(a) for the system in the presence of water diffusion no longer stay in one place in the presence of downhill water flow. They start moving from one side to another side [see Figs. \ref{Fig3}(b)-(d)], even diagonally, when the flow is in both $x$ and $y$ directions [similar to Fig. \ref{Fig4}], the Turing stability conditions are satisfied for each case. Sometimes, the water flow is insufficient to deform the Turing patch structure (e.g., the symmetric structure of the spots or labyrinthine); however, a good amount of water flow can force them to align in forming regular strips, and they move from one end to another. The moving speed and the distance between the two stripes depend on the water flow rate.

\begin{figure}[h]%
\centering
\includegraphics[width=0.9\textwidth]{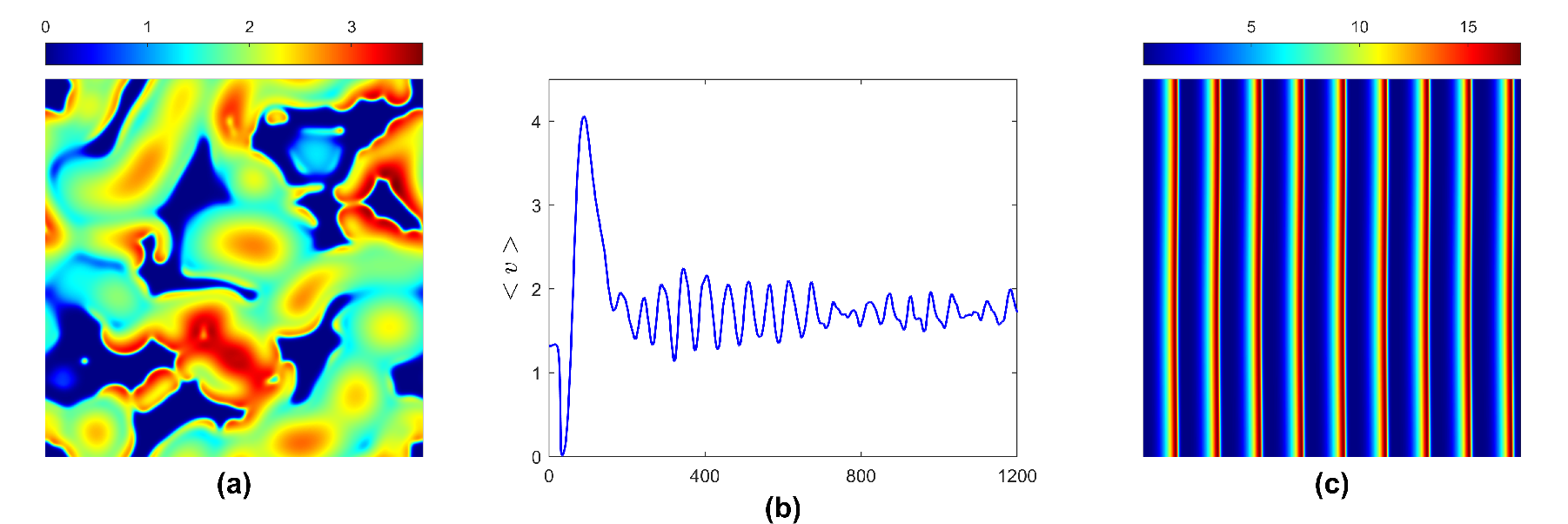}
\caption{ (Colour online) Non-homogeneous non-stationary solutions for the plant. Parameter values: $\alpha = 2.8125$, $\beta = 0.45$, $\gamma = 0.11$, $h = 1$, $\mu = 0.8$, $\eta = 0.05$, $a_{y} = 0$, $d_{u} = 2.5$, $d_{w} = 1.5$: (a) $a_{x} = 0$, (b) the spatial average of (a) in time, and (b) $a_{x} = 182.5$.}\label{Fig5}
\end{figure}

\subsubsection*{Time-varying patterns} 
Patterns may be stationary on a short-time scale, but they can be non-stationary over a longer time frame due to different pattern-forming mechanisms depending on the intensity of interaction and diffusivity. Sometimes, they oscillate regularly within a confined range or fluctuate irregularly, e.g., a regular moving stripe solution Fig. \ref{Fig3}(d). The regular case cannot arise without water flow due to the stability of the water-only state. However, irregular oscillatory solutions (chaos) and long transients can occur in a range of parameter values without groundwater flow in the region right to the magenta surface in Fig. \ref{Fig1}(a) [e.g., see Figs. \ref{Fig5}(a) and (b)]. In contrast, the temporal model predicts the existence of the water-only steady state. The diffusivity for water and herbivores has to be small for chaotic solutions so that they do not spread out as quickly, leading to stronger local interactions. Furthermore, these solutions are observed in a narrow range of diffusion parameter values, which cannot be observed without water diffusion. Here, both diffusion parameters have limits; beyond that, the system produces either non-homogeneous stationary patterns or goes for extinction. Nevertheless, these chaotic solutions transform into regular moving stripes in the presence of sufficient groundwater flow [see Fig. \ref{Fig5}(c)].

\subsubsection*{Catastrophic shift and prevention} 
As discussed earlier, the temporal model predicts catastrophic shifts for the parameter values in the Hopf unstable domain. This devastating shift can be seen with small diffusion coefficients and the absence of groundwater flow. For instance, the spatial model settles to the water-only steady state for $d_{u} = d_{w} = 1$. In addition, it produces chaos for some parametric combinations of the diffusion parameters in the absence of groundwater flow [see Fig. \ref{Fig5}(a)], and hence, slowly diffusing water can revive plants and herbivores from extinction. 

\begin{figure}[h]%
\centering
\includegraphics[width=0.52\textwidth]{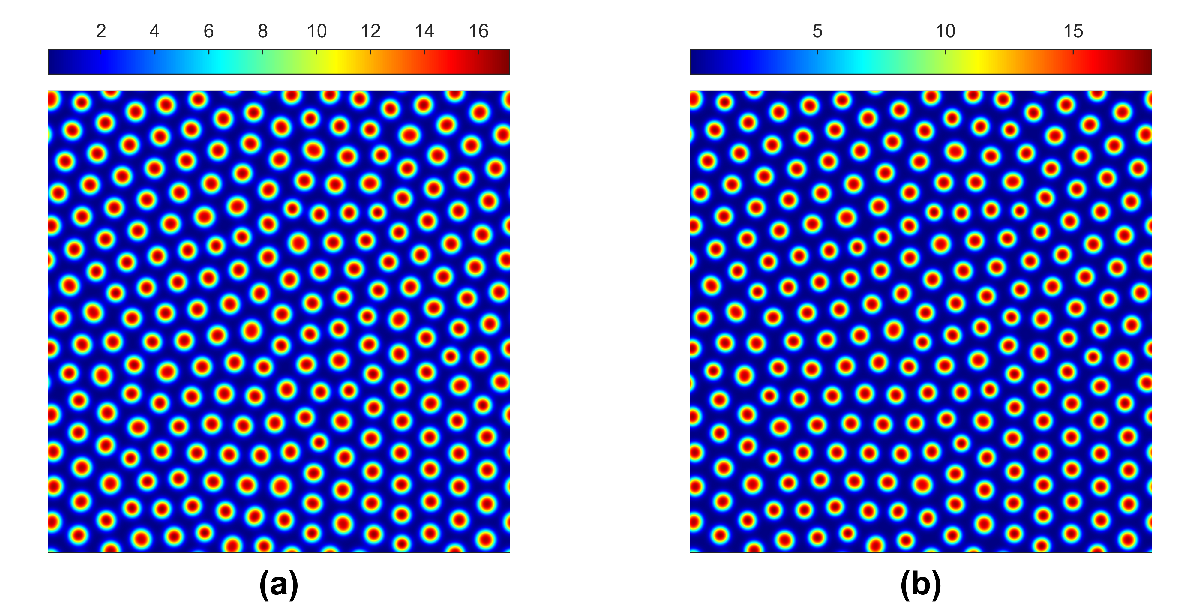}
\caption{(Colour online) Non-homogeneous non-stationary solutions corresponding to the second equation of (\ref{NDSTMV}). Parameter values: $\alpha = 2.8125$, $\beta = 0.45$, $h = 1$, $\mu = 0.8$, $\eta = 0.05$, $a_{x} = 0$, $a_{y} = 0$, $d_{u} = 500$, $d_{w} = 100$: (a) $\gamma = 0.1$ and (b) $\gamma = 0.11$ with (a) as the initial conditions.}\label{Fig6}
\end{figure}

The spatial model produces the Turing and non-homogeneous stationary patterns for higher diffusion rates for the water and herbivores (e.g., $d_{u} = 500$ and $d_{w}\geq 100$) on the right side of the blue surface in Fig. \ref{Fig1}(a) in the absence of groundwater flow. These patterns are robust because they remain stationary by shifting the background parameter value $\gamma$ left-to-right or right-to-left of the magenta surface in Fig. \ref{Fig1}(a). To verify these parametric shifts, we first simulate the spatial model till $t=2500$ for $\gamma = 0.1$ [see Fig. \ref{Fig6}(a)]. Then, the final non-homogeneous stationary solution is considered as the initial condition for $\gamma = 0.11$, and we rerun the simulation extended to $t = 3000$ [see Fig. \ref{Fig6}(b)]. We call this shifting forward shifting. In this case, the parameter values $\gamma = 0.1$ and $\gamma = 0.11$ are on the magenta surface's left and right in Fig. \ref{Fig1}(a). For this shifting of the parameter $\gamma$, the temporal model experiences catastrophic shifts; however, the spatial model leads to a coexistence scenario. On the other hand, the backward shifting of the parameter $\gamma$ from $0.11$ to $0.1$ also produces the same type of result. Furthermore, the system depicts the same behaviours for forward and backward shiftings for the parameters $\alpha$ and $\mu$.

The dynamics of the spatial model differ in this shifting for lower diffusion rates. The spatial model depicts the homogeneous stationary solution in the region left to the magenta surface in Fig. \ref{Fig1}(a) and either a chaotic solution or water-only stationary solution in the region right to the magenta surface in Fig. \ref{Fig1}(a) depending on the diffusion parameters. The forward shifting shows extinctions for the plants and herbivores. But the backward shift settles the system either to the coexisting steady state or the water-only steady state. In this case, the solution settles at the coexisting steady-state after the shifting when the system produces chaos before shifting. Otherwise, this shifting can not save the plants and herbivores from extinction. Furthermore, the stripe solutions can exist for such parameter values in the presence of water flow.

\section*{Conclusions}{\label{sec6}}

The construction of sustainable development in ecological civilization is a major concern nowadays. Ecosystems often face disruptions due to natural changes or external forces (e.g., climate, human intervention), particularly semiarid ecosystems. Researchers have been trying to understand these interferences using mathematical models and predict their effect on the ecosystem's functioning, and here is no exception. We have developed a mathematical model comprising water, plant, and herbivore interactions. Studying different parametric ranges enables us to find the vivid possible outcomes (e.g., coexistence, catastrophic shift, stationary patterns, long transients, etc.), for which it can capture the realistic scenarios of the organisms in semiarid ecosystems. The parameter values have been estimated based on the actual data.

The silico trials of the temporal model reveal several exciting results, such as herbivores' over-consumption of plants can cause vegetation collapse and dependencies (e.g., herbivores), excessive rainfall reduces plant growth, etc. These extinctions can be understood by the model described here in the region right to the magenta surface in Fig. \ref{Fig1}(a). The plant and herbivore can coexist in between the blue and magenta surfaces in Fig. \ref{Fig1}(a) with respect to all of the bifurcation parameters: water supply, plant consumption by the herbivores, and the conversion factor. In all three cases, restoring the coexisting state from extinction is impossible in the temporal setup, leading to a catastrophic shift. This catastrophic shift occurs due to the sub-critical Hopf bifurcation in the three-component model. No existing literature has identified the subcritical Hopf bifurcation behind such catastrophic change. Furthermore, the modified temporal model always has the water-only equilibrium point, which is linearly stable, and it does not have the coexisting equilibrium point for small parameter values of the bifurcation parameters (water-supply, consumption and conversion rates) [see Fig. \ref{Fig1}]. 

Vegetation abundance plays a pivotal role in semiarid ecosystems by influencing local climate conditions. An increment in the water supply ($\alpha$) helps coexist because agricultural production depends on water, which is observed in the considered model. Generally, the water assists positive plant growth, and more plant abundance supports the herbivores to survive. However, intensive rainfall can trigger critical transitions by destabilizing the system and altering the balance between vegetation growth, water availability and grazing by the herbivores, which leads to a significant yield loss. Furthermore, above a limit of water supply, depending on the other parameters, the ecosystem collapses. 

Semiarid ecosystems host diverse plant and animal communities that interact with each other and the environment. However, disruptions in these interactions can trigger critical transitions by destabilizing the ecological network \cite{sandacz2023}. This can be predicted through the considered model by increasing the consumption rate of the plants by the herbivores [see Fig. \ref{Fig1}]. The model shows that the herbivore species require a minimum food supply for survival. On the other hand, a higher consumption rate causes plant extinction due to over-exploitation, which further affects their dependencies, e.g., herbivores, and in the end, both species die [see Fig. \ref{Fig1}]. It also shows that a higher conversion rate ($\mu$) causes extinction, which proves that the herbivore cannot convert the maximum energy from the plant to power its life processes and to build more body tissues. This further validates that energy is lost whenever it is transferred from one trophic level to the next, called an energy pyramid, which decreases by going to higher trophic levels. 

In the last few decades, theoretical and empirical research has suggested that spatial heterogeneity strongly influences the distribution of organisms within their habitat and how populations colonize new areas and interact within a landscape \cite{schmitz2007}. The dynamics of a spatial model remain similar to the temporal model for lower diffusion rates but can be completely different for higher diffusion rates. We have shown different types of Turing stationary patterns in the absence of water flow in Figs. \ref{Fig2} and \ref{Fig3}(a). Furthermore, Fig. \ref{Fig2} shows that water diffusion is important in the resulting non-homogeneous distributions. 

In an ecosystem, the ground surface is also an essential factor. Water stands on flat ground, which primarily restricts plants' growth by blocking the movement of oxygen into the soil and carbon dioxide out of the soil. The system produces the Turing pattern only in the presence of water diffusion [see Fig. \ref{Fig3}(a)]. This shows groundwater diffusion is crucial for the plant population and overall system. Also, we have demonstrated that the Turing patterns can move from one spatial point to another in the presence of water flow [see Fig. \ref{Fig3}(b)], depending on the flow speed and its direction. An increase in the water flow speed can alter the geometry of patches. Still, in the end, they form regular strips moving from one end to another [see Fig. \ref{Fig3}(d)]. Therefore, these travelling strips are also Turing patterns but are not stationary. Thus, the Turing spatial structures further show the ecosystem's overall resilience.

The temporal model predicted excessive rainfall could destroy the plant and herbivore community. This extinction scenario can occur for the spatial model for higher rainfall without groundwater flow and low or zero water diffusion parameter values. But, for a higher diffusion, the system shows non-homogeneous stationary patterns, which are not Turing patterns \cite{turing1990}. These non-homogeneous stationary patterns can be sustained through forward and backward shifts as these parametric shifts can come at any time due to natural or human interventions. This further shows that the system is resilient in high water diffusion. Therefore, the spatial patterns can buffer the system against fluctuations, preventing rapid and catastrophic shifts. Furthermore, the water flow drives the system towards regular stripe patterns, which are more resilient, and it can be applied to other types of models where only diffusion is present. 

It is also true that diffusion can induce irregular distribution in a spatial model \cite{pascual1993, datseris2022}. In the considered model, the temporal counterpart suggests the extinction of plant and herbivore species beyond the Hopf bifurcation [see Fig. \ref{Fig1}]; however, the spatial model on the flat ground shows irregular time-varying distributions for some combinations of the diffusion parameters for the water and herbivores. This shows that self-organization allows the organisms' diversity and saves them from extinction. In addition, this chaos converted into regular strips moving from one end to another in the presence of water flow. Generally, low water diffusion occurs when the ground surface contains excessive rocks, and low diffusion occurs for herbivores when they are ill. Below these low diffusion parameters, plants and herbivores die; hence, the chaos introduced by the sub-critical Hopf bifurcation can be considered as the early warning signal of the catastrophic shift \cite{petrovskii1999, sherratt1997, banerjee2011}. Different regular patterns exhibit self-organizing properties that make them resilient to some environmental changes, while irregular patterns are more susceptible to catastrophic shifts under abrupt environmental conditions. 

Interestingly, the three-component spatio-temporal model proposed here can predict more realistic scenarios in an ecosystem than the currently existing literature. The silico trials show that the presence of herbivores in an ecosystem can cause catastrophic shifts; however, sufficient water diffusion or water flow can sustain the ecosystem in the long run. This shows that diversity can accelerate the extinction rate of species if we do not take precautions. Nevertheless, the modified model can be applied to the other types of herbivores, and in that case, the parameter values related to the herbivore species have to be calibrated. Additionally, our findings also set the stage for future studies exploring different natural and man-made factors in ecosystems, such as the spreading of fires, diseases, pollution, and introduction of invasive species \cite{touboul2018}.

\section*{Methods}{\label{sec2}}

Researchers have been working on a plant-water model in a semiarid environment \cite{klausmeier1999}, and it has been studied due to its simplicity and rich mathematical analysis. This spatio-temporal model is given by:
\begin{equation}{\label{STMV}}
    \begin{aligned}
    \frac{\partial u}{\partial t}&= a_{1} \frac{\partial u}{\partial x} + k_{1} - k_{2}u - k_{3}uv^{2},\\
         \frac{\partial v}{\partial t}&= d_{2} \bigg{(}\frac{\partial^{2}v}{\partial x^{2}} + \frac{\partial^{2}v}{\partial y^{2}}\bigg{)} + k_{3}k_{4}uv^{2} - k_{5}v,
    \end{aligned}
\end{equation}
where $u(x,y,t)$ and $v(x,y,t)$ denote the water and the plant densities, respectively. The uniform water supply rate is $k_{1}$, which is lost due to evaporation at the rate of $k_{2}u$. $k_{3}$ is the maximum rate of water consumption by plants, $k_{4}$ is the yield of plant biomass per unit of water consumed, and $k_{5}$ is the plant mortality rate. The parameter $a_{1}$ represents the water downhill speed, and $d_{2}$ is the plant dispersal diffusion rate. In this work, we introduce herbivore dynamics into the existing model (\ref{STMV}), and these herbivores fed the vegetation. To maintain the model's simplicity, the Holling type II functional response is considered with a constant death rate of the herbivore. Along with this, we also include water diffusion in the model \cite{kealy2012, siteur2014, eigentler2020}, which can capture the movement of surface water induced by spatial differences in infiltration rate \cite{rietkerk2002}. In addition, the water downfall in the $x$ and $y$ directions is considered separately in the model to capture a wide range of realistic scenarios. Including all these factors in Klausmeier's model \cite{klausmeier1999}, we obtain:
\begin{equation}{\label{NSTMV}}
    \begin{aligned}
         \frac{\partial u}{\partial t}&= d_{1} \bigg{(}\frac{\partial^{2}u}{\partial x^{2}} + \frac{\partial^{2}u}{\partial y^{2}}\bigg{)} + a_{1} \frac{\partial u}{\partial x} + a_{2}\frac{\partial u}{\partial y}+ k_{1} - k_{2}u - k_{3}uv^{2},\\
         \frac{\partial v}{\partial t}&= d_{2} \bigg{(}\frac{\partial^{2}v}{\partial x^{2}} + \frac{\partial^{2}v}{\partial y^{2}}\bigg{)} + k_{3}k_{4}uv^{2} - k_{5}v -\frac{k_{6}vw}{v+k_{7}},\\
         \frac{\partial w}{\partial t}&= d_{3} \bigg{(}\frac{\partial^{2}w}{\partial x^{2}} + \frac{\partial^{2}w}{\partial y^{2}}\bigg{)} + \frac{k_{8}k_{6}vw}{v+k_{7}}-k_{9}w,
    \end{aligned}
\end{equation}
where $u(x,y,t)$, $v(x,y,t)$ and $w(x,y,t)$ denote the densities of water, plants, and herbivores, respectively. Here, $d_{1}$ is the water diffusion, $a_{2}$ is the water flow in the $y$-direction, $k_{6}$ represents the rate of plant consumption by herbivores, $k_{7}$ is the half-saturation constant, $k_{8}$ is the conversion parameter where the plant biomass converts into herbivore biomass, and $k_{9}$ is the herbivore consumption rate at zero population growth. We non-dimensionalize this model by substituting $\widetilde{u} = uk_{4}\sqrt{k_{3}/k_{2}}$, $\widetilde{v} = v\sqrt{k_{3}/k_{2}}$, $\widetilde{w} = w\sqrt{k_{3}/k_{2}}$, $\widetilde{x} = x\sqrt{k_{2}/d_{2}}$, $\widetilde{y} = y\sqrt{k_{2}/d_{2}}$, and $\widetilde{t} = tk_{2}$ into (\ref{STMV}) and ignoring the tilde symbols, the model (\ref{NSTMV}) transforms into:
\begin{equation}{\label{NDSTMV}}
    \begin{aligned}
         \frac{\partial u}{\partial t}&= d_{u}\bigg{(}\frac{\partial^{2}u}{\partial x^{2}} + \frac{\partial^{2}u}{\partial y^{2}}\bigg{)} +a_{x} \frac{\partial u}{\partial x} +a_{y} \frac{\partial u}{\partial y} + \alpha - u - uv^{2},\\
         \frac{\partial v}{\partial t}&= \frac{\partial^{2}v}{\partial x^{2}} + \frac{\partial^{2}v}{\partial y^{2}} + uv^{2} - \beta v - \frac{\gamma vw}{v+h},\\
         \frac{\partial w}{\partial t}&= d_{w} \bigg{(}\frac{\partial^{2}w}{\partial x^{2}} + \frac{\partial^{2}w}{\partial y^{2}}\bigg{)} + \frac{\mu\gamma vw}{v+h}-\eta w,
    \end{aligned}
\end{equation}
where $\alpha = k_{1}k_{4}\sqrt{k_{3}/k_{2}^{3}}$, $a_{x} = a_{1}/\sqrt{k_{2}d_{2}}$, $a_{y} = a_{2}/\sqrt{k_{2}d_{2}}$, $\beta = k_{5}/k_{2}$, $\gamma = k_{6}/k_{2}$, $d_{u} = d_{1}/d_{2}$, $d_{w} = d_{3}/d_{2}$, $h = k_{7}\sqrt{k_{3}/k_{2}}$, $\mu = k_{8}$ and $\eta = k_{9}/k_{2}$. The habitat's area is large in an ecosystem; the periodic boundary conditions for all three variables are the most suitable in this case \cite{klausmeier1999}. All the numerical simulations are done for the model on a $200\times 200$ domain (in dimensional terms, 10,000 m$^2$) \cite{klausmeier1999}. The initial conditions for each of the simulations are considered as a heterogeneous perturbation around the homogeneous steady-state: $u_{0}(x,y) = u_{*} +\epsilon \xi_{ij}^{u}$, $v_{0}(x,y) = v_{*} +\epsilon \xi_{ij}^{v}$, and $w_{0}(x,y) = w_{*} +\epsilon \xi_{ij}^{w}$ with $\epsilon = 10^{-4}$, where $\xi_{ij}^{u}$, $\xi_{ij}^{v}$, and $\xi_{ij}^{w}$ are spatially uncorrelated Gaussian white noise terms.  

\textbf{Parameter choice}: The parameter values involved in the original model are given in \cite{klausmeier1999}. We consider the species ostriches as herbivores in the model. Ostriches feed the plant biomass; in captivity, they are typically fed between 3 and 4 lbs (1.3 kg to 1.8 kg) of food daily (https://birdfact.com/articles/what-do-ostriches-eat). In addition, for the spatial dimension, they need at least 1/3 of an acre (approximately 1349 m$^{2}$) per pair of birds. Here, we assume that each bird needs 1349 m$^{2}$ of spatial habitat to live, the same as a pair of birds. Therefore, $k_{6}$ = (1.3 to 1.8)$\times$365/1349 = 0.352 to 0.487 kg m$^{-2}$ individual$^{-1}$ year$^{-1}$. The mortality rate of baby ostriches is higher than that of juvenile ostriches. Combining both, the mortality rates of the ostriches range from 15\% to 50\% per year, i.e., $k_{9} = 0.15$ to $0.5$ year$^{-1}$.

\begin{table}[bt]
\caption{Parameter values for both dimensional and dimensionless models.}
\label{tab1}%
\begin{tabular}{clcl}
\hline 
Parameters & Value & Parameters & Value\\
\hline 
$k_{1}$  & $250$ to $750$ kg H$_{2}$O m$^{-2}$ year$^{-1}$ \cite{klausmeier1999} & $\alpha$ & 0.9375 to 2.8125 \\
$k_{2}$  & $4$ year$^{-1}$ \cite{klausmeier1999} & $\beta$ & 0.45 \\
$k_{3}$  & $100$ kg H$_{2}$O m$^{-2}$ year$^{-1}$ \cite{klausmeier1999} & $\gamma$ & 0.088 to 0.12175\\
$k_{4}$  & $0.003$ kg H$_{2}$O m$^{-2}$ year$^{-1}$  \cite{klausmeier1999}  & $h$ & 1.0 \\
$k_{5}$  & $1.8$ year$^{-1}$ \cite{klausmeier1999} & $\mu$ &  0 to 1 \\
$k_{6}$  & $0.352$ to $0.487$ kg individual$^{-1}$ year$^{-1}$ [est.] & $\eta$ & 0.0375 to 0.125 \\
$k_{7}$  & $0.2$ kg m$^{-2}$ [est.] & $a_{x}$ & 0 to 182.5  \\
$k_{8}$  & $0$ to $1$ [est.] & $a_{y}$ &  0 to 182.5  \\
$k_{9}$  & $0.15$ to $0.5$ year$^{-1}$ &  $d_{u}$ & 500 \\
$a_{1}$  & $0$ to $365$ m year$^{-1}$ \cite{klausmeier1999} & $d_{w}$ & $> 0$ \\
$a_{2}$  & $0$ to $365$ m year$^{-1}$ \cite{klausmeier1999} &   &  \\
$d_{1}$  & $500$ m$^{2}$ year$^{-1}$ \cite{deblauwe2008}& & \\
$d_{2}$  & $1$ m$^{2}$ year$^{-1}$ \cite{klausmeier1999} &   &  \\
$d_{3}$  & $> 0$ m$^{2}$ year$^{-1}$ &   &  \\
\hline  % Please only put a hline at the end of the table
\end{tabular}
\end{table}

\subsection*{Steady-states and their stabilities}
The equilibrium points for the water-plant model are the solutions of the nonlinear equations:
\begin{equation*}{\label{SEQ1}}
          \alpha - u - uv^{2} = 0,~\mbox{and}~uv^{2} - \beta v = 0.
\end{equation*}
This system has at most three solutions, and they are $(u_{0},v_{0})=(\alpha, 0)$, $(u_{1},v_{1}) = ((\alpha-\sqrt{\alpha^{2}-4\beta^{2}})/2,2\beta/(\alpha-\sqrt{\alpha^{2}-4\beta^{2}}))$, and $(u_{2},v_{2}) = ((\alpha+\sqrt{\alpha^{2}-4\beta^{2}})/2, 2\beta/(\alpha+\sqrt{\alpha^{2}-4\beta^{2}}))$. The feasibility condition for the last two equilibrium points is $\alpha > 2\beta$. Therefore, the water-plant model has three equilibrium points at most. The linear stability analysis shows $(u_{0},v_{0})$ and $(u_{1},v_{1})$ are stable, and $(u_{2},v_{2})$ is unstable. For studying Turing instability around the stable equilibrium point $(u_{1},v_{1})$, we substitute $u = u_{1}+\epsilon a_{1}\exp{(\lambda t +i(lx+my))}$ and $v = v_{1}+\epsilon b_{1}\exp{(\lambda t +i(lx+my))}$, where $\epsilon \ll 1$, into the non-dimensional version of the water-plant model and the linearization leads to
\begin{equation}{\label{SWPML}}
\begin{pmatrix}
    a_{11}-d_{u}(l^2+m^2)+i(la_{x}+ma_{y})-\lambda & a_{12}\\
    a_{21} & a_{22}-(l^2+m^2)-\lambda
\end{pmatrix}
\begin{pmatrix}
    a_{1}\\
    b_{1}
\end{pmatrix} = 
\begin{pmatrix}
    0\\
    0
\end{pmatrix},
\end{equation}
where $a_{11} = -1-v_{1}^{2}$, $a_{12} = -2u_{1}v_{1}$, $a_{21} = v_{1}^{2}$, and $a_{22} = 2u_{1}v_{1}-\beta$. In the absence of water flow (i.e., $a_{x} = a_{y} = 0$), the system  (\ref{SWPML}) reduces to 
\begin{equation*}{\label{SWPMLT1}}
\begin{pmatrix}
    a_{11}-d_{u}k^2-\lambda & a_{12}\\
    a_{21} & a_{22}-k^2-\lambda
\end{pmatrix}
\begin{pmatrix}
    a_{1}\\
    b_{1}
\end{pmatrix} = 
\begin{pmatrix}
    0\\
    0
\end{pmatrix},
\end{equation*}
where $k^{2} = l^2+m^2$. This system can have a non-trivial solution if the determinant of the coefficient matrix is equal to zero, which gives 
$$\lambda_{\pm}(k^2) = \frac{B(k^{2})\pm \sqrt{(B(k^{2}))^{2}-4C(k^{2})}}{2},$$
where $B(k^{2}) = a_{11}+a_{22}-k^{2}(d_{u}+1)$ and $C(k^{2}) = d_{u}k^{4}-(d_{u}a_{22}+a_{11})k^2+a_{11}a_{22}-a_{12}a_{21}$. As the equilibrium point $(u_{1},v_{1})$ is locally asymptotically stable for the temporal model, so $a_{11}+a_{22}<0$ and $a_{11}a_{22}-a_{12}a_{21}>0$, which implies $B(k^{2})<0$ for all $k>0$. Furthermore, $C(k^{2})>0$ for small positive $k$, but it can change its sign for a range of values of $k$ if $d_{u}>d_u^c$ holds \cite{murray2001}, where
$$d_u^c = \frac{a_{11}a_{22}-2a_{12}a_{21}\pm \sqrt{(a_{11}a_{22}-2a_{12}a_{21})^{2}-a_{11}^{2}a_{22}^{2}}}{a_{22}^{2}}.$$ 
Similarly, we find the equilibrium points for the system (\ref{NDSTMV}), which are the solutions of the nonlinear equations:
\begin{equation}{\label{SEQ}}
          \alpha - u - uv^{2} = 0,~~
          uv^{2} - \beta v -\frac{\gamma vw}{v+h} = 0,~\mbox{and}~\frac{\mu\gamma vw}{v+h}-\eta w = 0.
\end{equation}
Therefore, the model (\ref{NDSTMV}) always exists the water-only equilibrium point $(\alpha,0,0)$, and it is linearly stable. Also, it exists no-herbivore equilibrium points for $\alpha > 2\beta$, and they are $((\alpha+\sqrt{\alpha^{2}-4\beta^{2}})/2,2\beta/(\alpha+\sqrt{\alpha^{2}-4\beta^{2}}),0)$ and $((\alpha-\sqrt{\alpha^{2}-4\beta^{2}})/2,2\beta/(\alpha-\sqrt{\alpha^{2}-4\beta^{2}}),0)$. Furthermore, the system may have a unique coexisting equilibrium point $(u_{*},v_{*},w_{*})$, where 
$u_{*} = \alpha (\mu\gamma - \eta)^{2}/(h^2\eta^2 + (\mu\gamma - \eta)^{2})$, $v_{*} = h\eta/(\mu\gamma - \eta)$, and $w_{*} = \mu(\alpha - u_{*} - \beta v_{*})/\eta$. Here, the case $\mu\gamma = \eta$ is not true because we fix $h=1$. The Turing instability condition for this model can be studied following the same procedure as outlined for the water-plant model. 

\section*{Acknowledgments}

SP and RM thank the NSERC and the CRC Program for their support. RM also acknowledges the support of the BERC 2022--2025 program and the Spanish Ministry of Science, Innovation and Universities through the Agencia Estatal de Investigacion (AEI) BCAM Severo Ochoa excellence accreditation SEV-2017--0718 and the Basque Government fund AI in BCAM EXP. 2019/00432. This research was enabled in part by support provided by SHARCNET \url{(www.sharcnet.ca)} and Digital Research Alliance of Canada \url{(www.alliancecan.ca)}.

\section*{Declarations}

The authors declare no competing interests.

\bibliography{sample}% common bib file

%%\bibliography{sn-bibliography}% common bib file
%\input sn-article.bbl

\end{document}